\documentclass[aps,prl,reprint,superscriptaddress,amsmath,amssymb,a4paper,twocolumn,longbibliography]{revtex4-2}
 \usepackage{graphicx,graphics,color}

\usepackage{amssymb}
\usepackage{amsfonts}
\usepackage{epstopdf}
\usepackage{bm}
\usepackage{times,xspace}
\usepackage{color}
\usepackage[utf8]{inputenc}
\usepackage[normalem]{ulem}
\usepackage{xcolor}
\usepackage{multirow}
\usepackage{physics}
\usepackage{tensor}
\usepackage{siunitx}
\usepackage{lineno}
\usepackage{amsmath}
\usepackage{etoolbox} %% <- for \cspreto, \csappto

\usepackage[utf8]{inputenc}
\usepackage{amsmath}
\usepackage{amssymb,amsfonts,latexsym}
\usepackage{bm}
\usepackage[mathcal]{euscript}
\usepackage{graphicx}
\usepackage{epsfig}
\usepackage{color}
\usepackage{pifont,wasysym,marvosym}
\usepackage{textcomp}
\usepackage{comment}
\usepackage{epstopdf}
\usepackage{gensymb}
\usepackage{dcolumn}
\usepackage{hyperref}
\usepackage{physics}

\newcommand*\linenomathpatch[1]{%
  \cspreto{#1}{\linenomath}%
  \cspreto{#1*}{\linenomath}%
  \csappto{end#1}{\endlinenomath}%
  \csappto{end#1*}{\endlinenomath}%
}

\linenomathpatch{equation}
\linenomathpatch{gather}
\linenomathpatch{multline}
\linenomathpatch{align}
\linenomathpatch{alignat}
\linenomathpatch{flalign}

\begin{document}

\title{
%\cc{Work generating} limit cycles turn \cc{active} matter into robots\\
%Non-reciprocal transport in contracting vessels
Unidirectional flow from continuous broken symmetries%continuum Rectification
}

\author{Aaron Winn}
\thanks{These authors contributed equally}
\affiliation{Department of Physics and Astronomy, University of Pennsylvania, Philadelphia PA, 19104, USA}

\author{Justine Parmentier}
\thanks{These authors contributed equally}
\affiliation{Aix-Marseille Universit\'e, CNRS, Centrale M\'editerran\'ee, IRPHE, UMR 7342, 13384 Marseille, France}

\author{Eleni Katifori}
\affiliation{Department of Physics and Astronomy, University of Pennsylvania, Philadelphia PA, 19104, USA}
\affiliation{Center for Computational Biology, Flatiron Institute, New York, NY 10010, USA}

\author{Martin Brandenbourger}
\thanks{Corresponding author: martin.brandenbourger@univ-amu.fr}
\affiliation{Aix-Marseille Universit\'e, CNRS, Centrale M\'editerran\'ee, IRPHE, UMR 7342, 13384 Marseille, France}

% \affil[$*$]{\footnote{These authors contributed equally}}
% \affi

%\baselineskip24pt
%\baselineskip14pt
\maketitle
\medskip 
%\linenumbers
\textbf{
Locally broken symmetries are used across fields to transport matter, particles and information in preferential directions. Beyond local mechanisms, spatially distributed nonlinearities in crystalline media have enabled non-reciprocal transport, a rectification mechanism that operates continuously across scales and frequencies.
Here, we show that this concept applies beyond condensed matter, to fluid transport in living organisms and artificial systems. 
We take the example of the lymphatic vascular system, which transports interstitial fluid in mammals, and demonstrate that distributed leaflets act as continuous broken symmetries.
We build an artificial model of a collecting lymphatic and investigate the naturally richer dynamics of unidirectional transport that arises from spatiotemporal excitations. We observe robust and scalable transport for any waveshape and external pressure gradients. We show experimentally and theoretically that the contraction wavelength, directionality, and pulsatility control the flow rate. In particular, we counterintuitively find waveshapes that maximize transport when propagating against the direction of the flow.
Overall, our findings advance the understanding of unidirectional fluid transport in living systems and beyond, and reveal how coupling nonlinearities with spatiotemporal excitations can tune such transport across fields.
}

\section{Introduction}
\begin{figure}[t!]
%\hspace{-2.2cm}
\centering
%\hspace{0in}
\includegraphics[width=0.5\textwidth,trim=0cm 0cm 0cm 0cm]{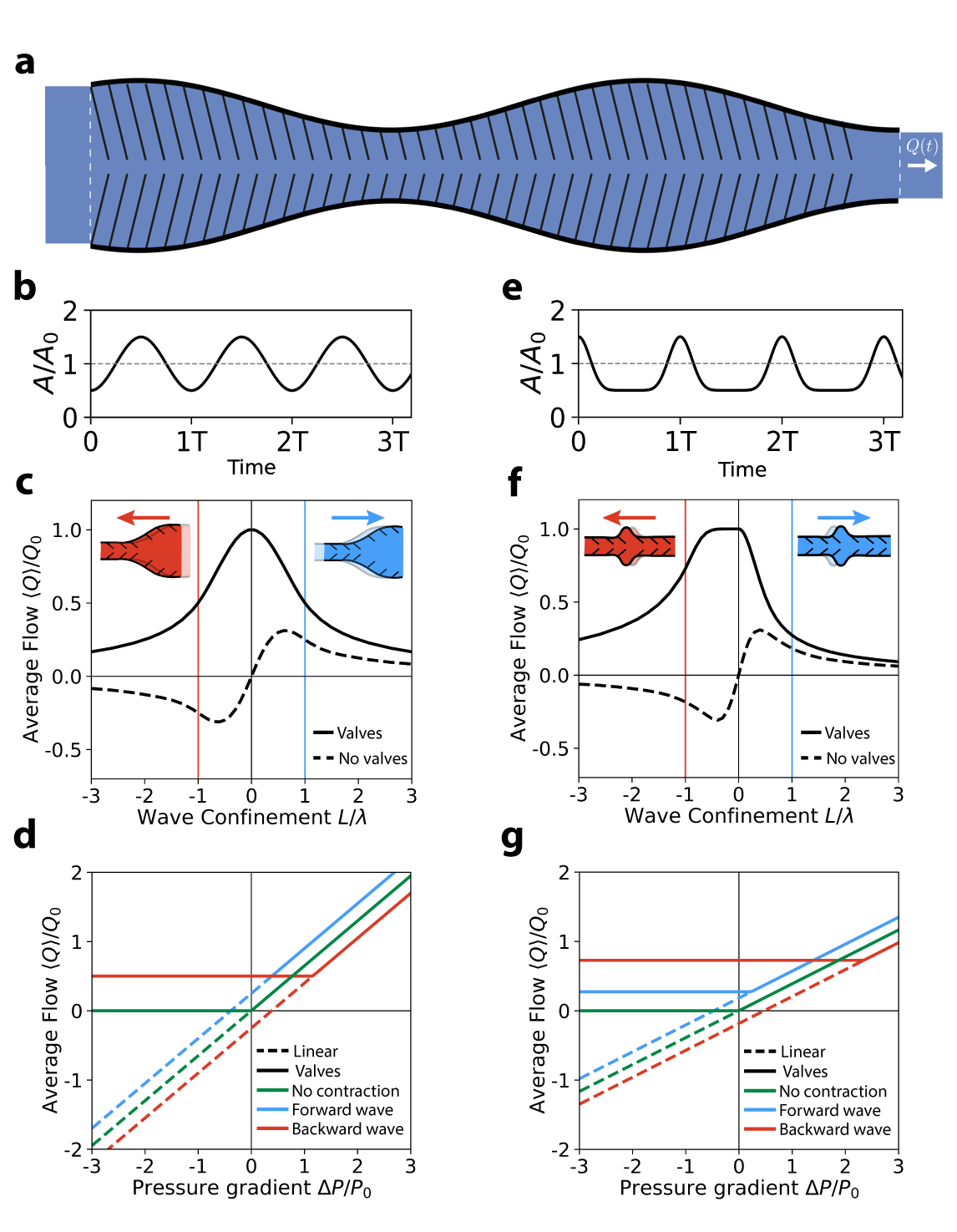}
\caption{\linespread{1.1}\selectfont{}
\textbf{Robust and scalable unidirectional flow in a contracting vessel with continuous valves.} 
\textbf{(a)} Schematic representation of the contracting vessel containing continuously distributed valves. 
\textbf{(b, c, d)} For sinusoidal contractions (\textbf{a}), valves in the vessel always impose positive flow rates (continuous lines) in opposition to vessels without valves (dashed lines). Robust flow rectification is observed independently of  the vessel length, the contraction wavelength (\textbf{c}), or the pressure gradient (\textbf{d}). $\Delta P$ is defined as the pressure at the inlet (left end) minus the pressure at the outlet (right end), meaning that a negative value of $\Delta P$ corresponds to an adverse pressure gradient. The directionality of the wave propagation has no influence on the flow rate until large favorable pressure gradients.
\textbf{(e, f, g)} For pulsatile contractions (\textbf{e}), the vessel retains the previous properties, but optimal transport is surprisingly observed for backward propagating waves until large favorable pressure gradients. 
}
\label{Fig1}
\end{figure}

Rectification, the ability to convert oscillating excitations into directional transport is a hallmark of broken symmetry in physics, from condensed matter \cite{erickson2007fundamentals} to fluid mechanics and biology, most notably with regard to fluidic pumping\cite{alvarado2017nonlinear,mosadegh2010integrated,10.1039/9781782628491,wehner2016integrated,park2018viscous,martinez2024fluidic,winn2024operating,mosadegh2010integrated,10.1039/9781782628491,park2018viscous,brandenbourger2020tunable,nain2024tunable,rothemund2018soft,nguyen2021flow}. In most cases, rectification has been understood as arising from localized nonlinearities, such as diodes in electronics or valves in fluidics \cite{moore2018lymphatic,lu2018biaxial,levin2024asymmetric,park2021fluid}, that impose directional bias at discrete points in a system.  Recent research in condensed matter has broken new ground by extending rectification to continuous systems \cite{tokura2018nonreciprocal,ideue2021symmetry,kim2024intrinsic,du2021nonlinear,nakamura2017shift,xiao2020berry,sollner2015deterministic,isobe2020high,zhang2021terahertz}. This has enabled scalable directional transport, called non-reciprocal transport, that remains robust against excitation across the material, with broad implications from energy harvesting  \cite{du2021nonlinear,nakamura2017shift,isobe2020high} to quantum information processing  \cite{ideue2021symmetry,du2021nonlinear,sollner2015deterministic}. %Such continuum rectification can emerge from nonlinearities distributed across crystalline structures \cite{tokura2018nonreciprocal,ideue2021symmetry,kim2024intrinsic,du2021nonlinear,nakamura2017shift,xiao2020berry,sollner2015deterministic,isobe2020high,zhang2021terahertz}. This is the case for the Nonlinear Hall Effect, characterized by a Hall voltage nonlinearly dependent on perpendicular driving currents. 
In biology, some vascular networks in plants and mammals \cite{moore2018lymphatic,park2021fluid,lu2018biaxial,levin2024asymmetric} incorporate distributed nonlinearities along fluidic networks to regulate directional flow. These vascular systems often exhibit rich spatiotemporal dynamics that extend beyond a single valve, suggesting that the combination of distributed nonlinearities and spatiotemporal dynamics can underlie collective rectification, akin to the emerging frameworks of non-reciprocal transport.
%Previous studies have characterized how individual valves enable directional control in a vessel \cite{...}. However, many biological vascular systems exhibit rich spatiotemporal dynamics, such as propagating vessel contractions. 
\begin{figure*}[!t]
%\hspace{-2.2cm}
\centering
\hspace{0in}
\includegraphics[width=\textwidth,trim=0cm 0cm 0cm 0cm]{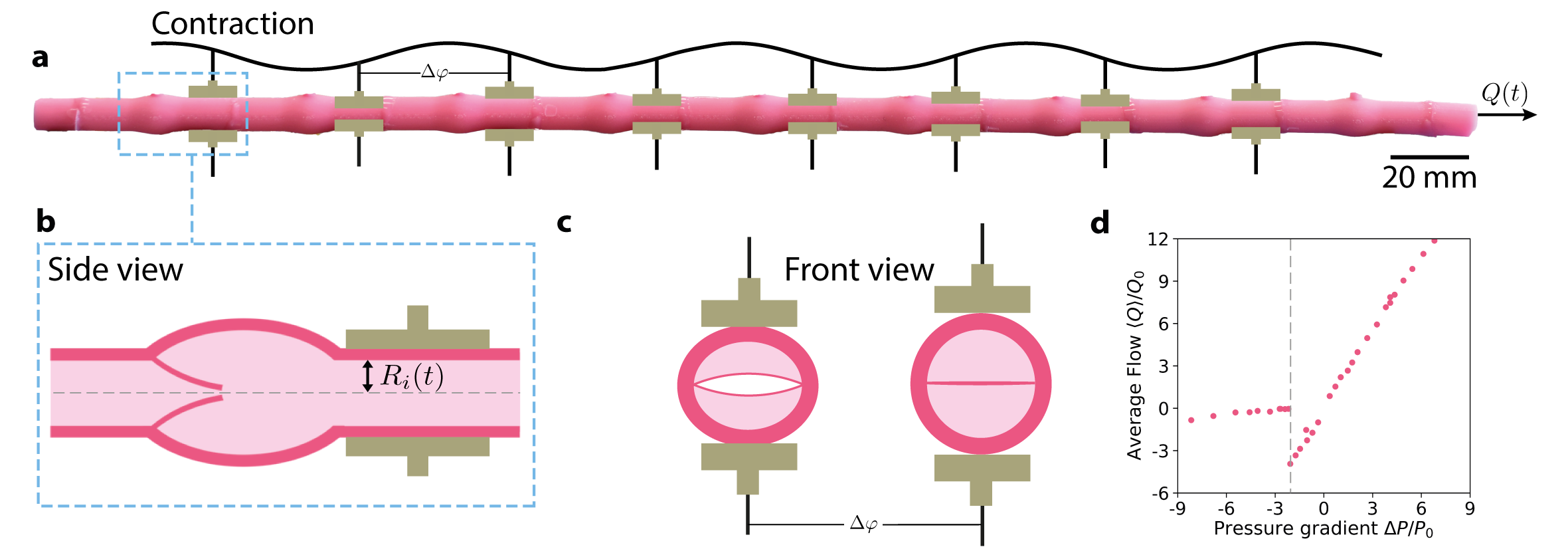}
\caption{\linespread{1.1}\selectfont{}
\textbf{Contracting vessel with distributed valves} \textbf{(a)} Artificial lymphatic vessel design inspired by imaging of rat lymphatics \cite{bohlen2009phasic,leak1980lymphatic}. \textbf{(b,c)} Sketch of the vessel cross-section. The vessel consists of repeating segments called lymphangions, separated by valves. An actuator locally contracts each lymphangion with an amplitude $R_i(t)$, where $i$ indicates the actuator index.  \textbf{(d)} Flow-pressure response of a single valve. For large adverse pressure gradients, the valve is closed and no flow is measured. For favorable pressure gradients, the valve is open and the flow linearly increases with $\Delta P$. Contrary to perfect valves described in Fig. \ref{Fig1}, a negative flow rate is measured for small adverse pressure gradients. }
\label{Fig2}
\end{figure*}
%Studies have characterized how individual valves enable directional flow control in a vessel \cite{mosadegh2010integrated,10.1039/9781782628491,park2018viscous,brandenbourger2020tunable,nain2024tunable,rothemund2018soft}. Yet, many biological vascular systems exhibit rich spatiotemporal dynamics beyond one valve, suggesting that the combination of distributed nonlinearities and spatiotemporal dynamics might underlie more robust and scalable unidirectional transport.
This raises two questions: Can such fluid transport be understood as non-reciprocal, meaning that unidirectional flow emerges reliably from arbitrary excitations and across scales? If so, can spatiotemporal actuation, beyond the temporal driving studied in condensed matter, unlock distinct transport regimes?

Here, we draw inspiration from the lymphatic vascular system, where distributed valves and coordinated contractions drive interstitial fluid transport. We first model how imposed contractions in a vessel with continuously distributed valves generate unidirectional transport that is robust to waveshape, vessel length, and external pressure. We also uncover regimes in which non-reciprocal transport is paradoxically enhanced by excitations propagating against the flow. Finally, we create an artificial vessel reproducing lymphatic geometry and show, experimentally and theoretically, that the same behavior arises in a soft tube containing a finite but sufficiently dense set of leaflets.

\section{Results}

We first investigate theoretically the emergence of robust and scalable unidirectional fluid transport in a vessel with continuously distributed valves (Fig. \ref{Fig1}a). We consider a cylindrical vessel filled with a Newtonian fluid and driven by an imposed area $A(x,t)$. The contraction is assumed axisymmetric with radius $R(x,t)$ such that $A(x,t) = \pi R(x,t)^2$. By integrating over the cross-section, the fluid dynamics within the vessel can be completely characterized by the pressure $P(x,t)$ and flow $Q(x,t)$. We consider an incompressible fluid at low Reynolds number under the lubrication approximation. The continuity and momentum equations in this problem read 
\begin{equation}
    \frac{\partial Q}{\partial x} + \frac{\partial A}{\partial t} = 0
\end{equation}
\begin{equation}
    \frac{\partial P}{\partial x} + \frac{8\pi \mu}{A^2} Q = 0.
\end{equation}
We consider a finite pipe of fixed length $L$ and rest area $A_0$ with no valves. The pipe is actuated by a traveling wave such that the area varies as $A(x,t) = A_0[1 - \eta_A \cos{(2\pi (x/\lambda-t/T))}]$ (Fig. \ref{Fig1}b). We study the time averaged outflow $\langle Q \rangle$, normalized by  $Q_0=\frac{A_0 L}{T}$. We recover the classic result of peristaltic waves (Fig. \ref{Fig1}cd, dashed lines) \cite{Shapiro1969PeristalticNumber, Li1993Non-SteadyTubes}. Synchronous contractions ($\lambda=\infty$) do not generate fluid transport (Fig. \ref{Fig1}c, dashed lines). For propagating waves, the amplitude and directionality of the average flow are set by the ratio of the vessel length and the contraction wavelength $L/\lambda$ (Fig. \ref{Fig1}c, dashed lines), and the pressure gradient across the vessel $\Delta P$ (Fig. \ref{Fig1}d). A positive (negative) value of $L / \lambda$ corresponds to a forward (backward) propagating wave. Forward (backward) propagating waves and favorable (adverse) pressure gradients contribute to positive (negative) flow rates. The scaling of the average flow with the pressure gradient is set by the average area of the vessel $\langle A \rangle$.

%For stationary contraction waves, the slope of the linear increase is set by the average resistance of the vessel. For propagating contraction waves, the average flow increases or decreases depending on their orientation (Fig. \ref{Fig1}d). 
%By considering finite pipes of constant length $L$ and travelling waves $A(x,t) = A(x-ct)$, we recover the classic result of peristaltic waves (Fig. \ref{Fig1}c, dashed lines). 
%The Fig. \ref{Fig1}c.
%The contraction wave increases or decreases the average flow $< Q >$ depending on its orientation. However, 
These observations critically fail when a non-linear pressure-flow relationship is taken into account. To illustrate this, we consider a vessel containing a continuum of ideal valves that remain closed and only open to permit forward flow (Fig. \ref{Fig1}a). Building on earlier analyses of infinitely long tubes \cite{winn2024operating}, we demonstrate (see Supplementary Information Section 2 \& 3) that for a tube of finite length and imposed area deformation, at most one valve among the continuum remains closed. We show that for any traveling wave $A(x,t) = A(x-ct)$, the flow reads
\begin{equation}\label{eq2}
    Q(x,t) = cA(x,t) - (cA)_{\text{min}}(t) + Q^{\text{nv}}_{\text{min}}(t) \Theta(Q^{\text{nv}}_{\text{min}}(t)).
\end{equation}
where $A_{\text{min}} (t)$ is the minimum area of the vessel and $Q^{\text{nv}}_{\text{min}}(t) \Theta(Q^{\text{nv}}_{\text{min}}(t))$ indicates that the flow cannot exceed that of the non-valve case (with $Q^{\text{nv}}_{\text{min}}(t)$ the minimum flow in the valveless system). 
For synchronous contractions, Eq. \ref{eq2} demonstrates flow rectification across all contraction frequencies (See Supplementary Information Section 3), illustrating how principles of non-reciprocal transport, previously established in condensed matter physics \cite{tokura2018nonreciprocal,ideue2021symmetry,kim2024intrinsic,du2021nonlinear,nakamura2017shift,xiao2020berry,sollner2015deterministic,isobe2020high,zhang2021terahertz}, reemerge within the context of fluid mechanics.
Building on the insights from synchronous contractions, we investigate propagating waves and observe that the direction of the contraction wave does not matter anymore in the presence of distributed asymmetries. For any wave orientation (Fig. \ref{Fig1}c, continuous lines) the flow is rectified forward ($\langle Q \rangle > 0$). This observation is maintained for any pressure gradient (Fig. \ref{Fig1}d, with $P_0=\mathcal{R} Q_0$ and $\mathcal{R}$ the hydraulic resistance). The forward flow rectification is expected in the presence of ideal valves, however,  counterintuitively, up until large favorable pressure gradients, the average flow rate is perfectly symmetric for contractions propagating forward or backward. This demonstrates that vessels containing distributed leaflets generate unidirectional fluid transport independently of their size (scalability) and spatiotemporal excitation (robustness). Overall, despite behaving almost everywhere like a Newtonian, viscous fluid which obeys the reciprocal theorem \cite{MasoudStone2019}, the presence of valves introduces new surface stresses that lead to such non-reciprocal transport (see Supplementary information Section 4).

While fluid transport is always rectified in the direction of the asymmetry, its amplitude is set by the contraction applied to the vessel. Eq. \ref{eq2} shows that the unidirectional flow is directly connected to the symmetry of the contraction wave via the term $(cA)_{\text{min}}(t)$. To illustrate this, we consider a pulsating contraction wave $A(x,t) = A_0 [1.5 - \eta_A \cos^8{\pi (x/\lambda-t/T)}]$, for which the vessel is more often contracted than relaxed  (Fig. \ref{Fig1}e). If the vessel contains no valves (Fig. \ref{Fig1}fg, dashed lines), we simply recover the previous results with an overall decrease in the amplitude of the flow rate. Such a decrease is explained by the reduction of the average area of the vessel during a contraction $\langle A \rangle $. If the vessel contains continuously distributed valves, the flow is once again rectified independently of the vessel length or the contraction wavelength (Fig. \ref{Fig1}f), and the imposed pressure gradient (Fig. \ref{Fig1}g). By contrast, the pulsating wave shown in Fig. \ref{Fig1}e leads to unidirectional flow with an unprecedented feature, namely larger unidirectional transport occurs for contractions propagating against the average flow. This is solely set by the asymmetry between vessel contraction and relaxation. We show in the Methods that pulsating contractions with the opposite asymmetry lead to a larger flow rate for forward propagating waves (Extended Data Fig. \ref{FigSI1}). We extend these results to any waveshape in the Supplementary Information (Section 3) and the Methods and show that the localization of the contraction increases the difference between backward and forward propagating waves.  Beyond scalability and robustness, these results demonstrate that the waveshape can be designed to optimize flow rectification in any regime of interest.

So far, the results shown in Fig. \ref{Fig1} remain predictions from a theoretical model with significant assumptions. Does the prediction of non-reciprocal transport, which implies unidirectional flow for arbitrary excitations across scales, hold in experimental settings where a continuous valve distribution cannot be implemented? To answer this question, we draw inspiration from the lymphatic vascular system, known for its robust unidirectional fluid transport. In the lymphatic system, the collecting lymphatics are vessels transporting interstitial fluid across the body of mammals \cite{moore2018lymphatic}. They are subdivided into short segments called lymphangions by regularly spaced valves. Lymphangions have a diameter between 80 to 2800 $\mu m$ and are embedded with smooth muscles enabling contractions at a rate varying between 0.1 to 0.35 Hz. The combination of valves and organized or disorganized contractions has been observed to generate robust fluid transport against varying pressure inputs \cite{moore2018lymphatic}. Such unidirectional fluid transport occurs at low Reynolds numbers with peristaltic waves that have been observed propagating both along and against the flow direction.

\begin{figure}[t!]
%\hspace{-2.2cm}
\centering
\hspace{0in}
\includegraphics[width=0.5\textwidth,trim=0cm 0cm 0cm 0cm]{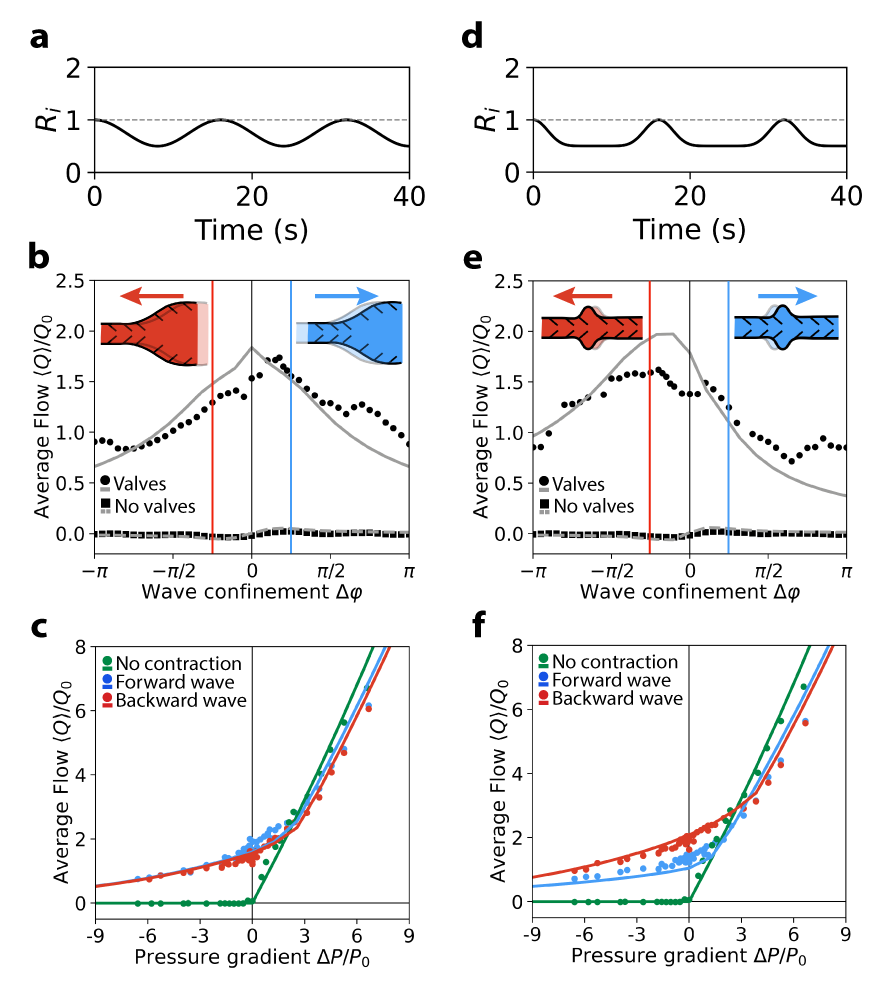}
\caption{\linespread{1.1}\selectfont{}
%\textbf{Scalable and robust unidirectional flow regimes} 
\textbf{Unidirectional Flow from distributed broken symmetries in a soft vessel: experiment and theory.} 
\textbf{(a, b, c)} For sinusoidal contractions (\textbf{a}), valves in the vessel always impose positive flow rates (continuous lines) in opposition to vessels without valves (dashed lines). Robust flow rectification is observed independently of the imposed pressure gradient (\textbf{c}), the vessel length or the contraction wavelength (\textbf{b}). The directionality of the wave propagation has no influence on the flow rate until large favorable pressure gradients.
\textbf{(d, e, f)} For pulsatile contractions (\textbf{d}), the vessel retains the previous properties, but optimal transport is surprisingly observed for backward propagating waves until large favorable pressure gradients. The continuous lines correspond to the theoretical model considering soft vessels.
}
%The reported flow rate is normalised by $Q_0=\frac{2 \Delta\!V}{T}= 7.5$ ml/min, where $\Delta V$ is the contracted volume of the vessel and $T$ is the contraction period.
\label{Fig3}
\end{figure}

Here, we build a model of a collecting lymphatic vessel made of 8 lymphangions (9 valves) by matching images taken from rat lymphatics \cite{bohlen2009phasic,leak1980lymphatic} (Fig. \ref{Fig2}a). The artificial lymphangion is upscaled by a factor 20. It is made of elastomer (see methods) and has a minimal inner radius of $r_1=4$ mm and a maximal radius of $r_2=5.5$ mm in the sinus region, with a wall thickness of 3 mm across the vessel. The vessel is filled with a mix of \textit{UCON} oil and water of viscosity $\mu =$ 108 cSt to ensure flow at low Reynolds number $Re=\rho Q/(\pi r \mu)\sim10^{-1}$. Each lymphangion is separated by a two-leaflet valve (Fig. \ref{Fig2}b). The valve acts as a diode: it closes (opens) in response to adverse (favorable) pressure gradients. In practice, the valve departs from the ideal diode, remaining open under small adverse pressure gradients (Fig. \ref{Fig2}d). The artificial lymphangions are contracted by servomotors (Dynamixel AX-12A) connected to a rack-pinion-rack assembly that converts rotation into linear, symmetric contraction of the lymphangions (Supplementary Information Section 6).  Each servomotor $i$ is placed at regular intervals. Each induces a lateral compression (Fig. \ref{Fig2}b) of the form:

\begin{equation}\label{eqXP}
    R_i(t) = 1-\frac{\eta}{\alpha} [1-\cos^\beta{(\frac{\alpha\pi t}{T} -i \,{\Delta\varphi})}],
\end{equation}
where $R_i(t)$ is the horizontal radius of the vessel at the contraction point of the servomotor $i$ nondimensionalized by its initial radius $r_1$, $\eta =0.25$ is the contraction amplitude, $T=16$ s is the contraction period, $\Delta \varphi \in [-\pi,\pi]$ is the phase difference between each contraction and characterizes the wave confinement. The contraction amplitude $\eta$ was selected sufficiently large to ensure that the non-ideal characteristics of the diode remained negligible. Although the vessel deformation is expected to be more complex than that in Fig. \ref{Fig1}, the framework allows us to characterize the system's response to sinusoidal ($\alpha =2,\beta=1$, Fig. \ref{Fig3}a) and pulsatile ($\alpha =1, \beta=8$, Fig. \ref{Fig3}d) contractions.
%The local contraction of each servomotor generates a contraction wave with a wavelength $\lambda= \frac{d}{\varphi \pi}$, where $d$ is the distance between each motor.

To test the predictions of the valve continuum model, we investigate experimentally fluid transport emerging from contractions of the artificial lymphatic vessel. 
%We define $\langle Q \rangle = \frac{\langle Q' \rangle}{Q_0}$, where  $Q_0=\frac{2 \Delta\!V}{T}= 7.5$ ml/min with $\Delta V$ the contracted volume of the vessel and $T$ the contraction period. 
We first investigate fluid transport for synchronous contractions and recover flow rectification for any contraction frequency (See Supplementary Information and Extended Fig. \ref{FigSI2}). The results are normalized by the characteristic flow rate $Q_0=\frac{2 \Delta\!V}{T}= 7.5$ ml/min with $\Delta V$ the contracted volume of the vessel and $T$ the contraction period. For propagating contractions, we write $\lambda= \frac{2 \pi d}{\Delta \varphi }$, such that $\Delta \varphi$ directly relates to the wave confinement $L/\lambda$. We observe flow rectification for any wave confinement (Fig. \ref{Fig3}b) in the presence of valves. The flow rate is also slightly asymmetric, with a maximum flow observed for forward propagating waves. Fig. \ref{Fig3}c shows a rectified flow for any pressure gradient independently of the directionality of the excitation (experiments performed at $\Delta \varphi=0.9$). The applied pressure gradient is normalized by the characteristic pressure $P_0=\mathcal{R} Q_0 = 18.4$ mbar, where $\mathcal{R}$ is the hydraulic resistance of the vessel. The results in Fig. \ref{Fig3} are similar to the observation in Fig. \ref{Fig1}. Yet, some striking differences appear between Fig. \ref{Fig1} and Fig. \ref{Fig3}. For example, the rectified flow rate is not constant for adverse pressure gradient. Furthermore, vessel contractions appear to induce only minimal flow in the absence of valves.

We show that the differences between Fig. 1 and Fig. 3 are due to the vessel compliance by formulating a theoretical model considering contractions of a compliant vessel characterized by a resistance $\mathcal{R}$ and a compliance $\mathcal{C}$ (see Supplementary information Section 5). We consider a continuous contraction force $P_{ext}$, such that the vessel volume deformation is expressed as $A(x,t)L-A_0 L = \mathcal{C}[P(x,t)-P_{ext}(x,t)]$. This force mimics the experimental configuration, in which the vessel radial deformation is a combination of local contractions and vessel relaxation. The relaxation time of the vessel is $ \tau = \mathcal{RC}/2$. By applying a sudden contraction to the tube, an experimental value of $\tau = 10.95$ s (see Supplementary information Section 6), comparable to the driving period $T = 16$ s, was found. Using this experimentally obtained relaxation time and the measured total vessel resistance $\mathcal{R}$, the model for force-imposed peristalsis with valves has only a single fitting parameter: the driving amplitude $\eta_P$, defined as the amplitude of the applied pressure normalized by $A_0 L/T$. Using $\eta_P = 0.18$, the model captures well the evolution of the flow rate with the wave confinement and the pressure gradient with and without valves. It also captures the effect of pulsatile contractions (Fig. 3ef), successfully recovering the regime in which backward-propagating waves optimize forward transport. We show that the same conclusions apply for different pulsating signals (Extended Fig. \ref{FigSI1}). Crucially, the soft vessel model demonstrates that continuous rectification is achievable through a discrete arrangement of leaflets, as long as the distance between leaflets $d$ does not exceed the damping length scale, $\sqrt{\frac{L^2 T }{\mathcal{RC}}}$.

Overall, our results show that despite the shape of the vessel, its compliance, the imperfection of the valves, the discrete actuation and the limited number of valves, we recover the different modes of non-reciprocal transport described by Fig. \ref{Fig1}, underscoring the robustness conferred by distributed valves.

%For symmetric contractions (Fig. \ref{Fig3}a), we observe a rectified flow for any wave confinement $\Delta \varphi$ (Fig. \ref{Fig3}b). In the continuous contraction limit, the wave confinement can be written as %$\phi=\frac{2 \pi L}{\lambda}$
%$\lambda= \frac{2 \pi d}{\Delta \varphi }$
%, which indicates that the experiments follow the same trend as in \ref{Fig1}c. Note that only a small average flow is induced in a channel containing no valves (square symbols in Fig. \ref{Fig3}b) because its overall flow resistance drops (Supplementary information Fig. 2). 

\section{Conclusion and Outlook}

We have shown how distributed asymmetries in vessels enable robust and scalable unidirectional transport against any pressure gradient and for any spatiotemporal actuation.
%Our experimental and theoretical results provide a quantitative explanation of how leaflets in contracting vessels drive robust transport across a large area. 
This framework helps explain similarly robust flow behaviors observed in biological systems such as the lymphatic vasculature, features previously reported but not understood \cite{moore2018lymphatic}. These results contribute to a growing body of work exploring how distributed nonlinearities control fluid flows \cite{alvarado2017nonlinear,garcia2025spontaneous,wehner2016integrated,martinez2024fluidic,paludan2024elastohydrodynamic}. We envision that these mechanisms will provide new vistas for applications that rely on unidirectional transport, e.g., pumping in microfluidics \cite{iakovlev2022novel,cacucciolo2019stretchable} and wearable fluidics \cite{smith2023fiber}, crawling robots\cite{tirado2024earthworm}, particle transport \cite{reichhardt2017ratchet}, and energy harvesting \cite{van2007power}. We expect that our characterisation of new transport regimes emerging from spatiotemporal driving will stimulate further research on non-reciprocal transport beyond fluid mechanics \cite{tokura2018nonreciprocal,ideue2021symmetry,kim2024intrinsic,du2021nonlinear,nakamura2017shift,xiao2020berry,sollner2015deterministic,isobe2020high,zhang2021terahertz}.

\textbf{Acknowledgments.} We thank William Le Coz and Eric Bertrand for technical assistance and Juan Huaroto Sevilla and Pierre Lambert for valuable discussions. M.B. and J.P. acknowledge funding from the European Research Council under Grant Agreement No.~101117080. E.K. and A.W. acknowledge funding from  the Army Research Office (ARO) through the Multidisciplinary University Initiative (MURI) Grant No. W911NF2210219, and the University of Pennsylvania Materials Research Science and Engineering Center (MRSEC) through Grant No. DMR-2309043.

\textbf{Author contributions.} A.W., J.P., E.K., and M.B. designed the research. J.P. and M.B. designed the experiments and performed the measurements. A.W., E.K. and M.B. performed the numerical simulations. A.W. and E.K. carried out the theoretical calculations. All authors contributed to the interpretation of the data and wrote the paper.

\section{Methods}

\setcounter{figure}{0}
\renewcommand{\figurename}{Extended Data FIG.}

\subsection{Realization of the vessel with distributed broken symmetries}

The vessel shown in Fig. \ref{Fig2}a of the main text consists of 8 lymphangions modeled from images taken from rat lymphatic vessels \cite{bohlen2009phasic,leak1980lymphatic}. The vessel is built by moulding sections of the vessel containing one valve. These sections are made of Polyvinyl siloxane (Zhermack Elite Double 8, Young's modulus of 186 kPa \cite{lazarus2015youngModulus}). The same Polyvinyl siloxane is used to connect each section by reticulation. The vessel has an inner minimal radius $r_1 = 4$ mm and maximal radius $r_2 = 5.5$ mm, a vessel thickness $h = $ 3 mm and a total length $L = 620$ mm (see Fig. \ref{Fig2}a of the main text). Each valve corresponds to a pair of symmetric leaflets, of length 10 mm and thickness of about 1 mm, separated at rest by a maximum aperture of 0.5 mm. 
%The vessel is filled with a mixture of Ucon oil and water, having a viscosity $\mu = \SI{108}{\milli\pascal\second}$ and a density \SI{1060}{\kilo\gram\per\cubic\meter}. 
Two vessel sections are separated by 62 mm.

The vessel is actuated by 8 servomotors (Dynamixel AX-12A). Each servomotor angular displacement actuates a rack-pinion-rack assembly (Shown in the Supplementary Information Section 6) with a precision of 0.3 \textdegree. This leads to a symmetric contraction of the vessel. This contraction takes place 10 mm after the end of the valves on a width of 20 mm. The servomotors are interfaced with a custom Python script.

\subsection{Characterization of the non-reciprocal fluid transport}

The flow rate was measured by two Sensirion flow meters (model SLF3S-4000B) placed at each end of the vessel. Each flow meter was connected to an external container. The pressure gradient was imposed by either changing the liquid height in the upstream or downstream container (up to 30 mbar), or using an Elveflow pressure controller (from 30 to 125 mbar, model OB1 MK3+). The data shown in Fig. \ref{Fig3}cf of the Main text was adjusted to incorporate the effects of linear hydraulic resistance associated with the external connections.

\subsection{Extended Regimes of Non-Reciprocal Fluid Transport}

\begin{figure}[t!]
%\hspace{-2.2cm}
\centering 
\hspace{0in}
\includegraphics[width=0.3\textwidth,trim=0cm 0cm 0cm 0cm]{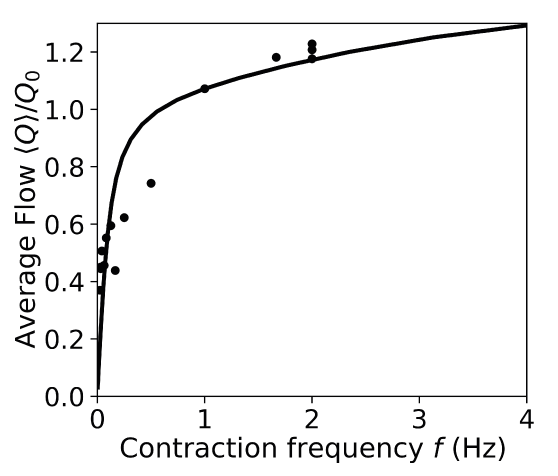}
\caption{\linespread{1.1}\selectfont{}
\textbf{Flow rectification for synchronous contractions} The black dots correspond to measurements and the continuous lines to the model obtained with a sinusoidal contraction. In both cases, the flow rate is expressed in dimensionless form using the characteristic flow rate $Q_0$, defined at the contraction frequency introduced in the main text.
}
\label{FigSI2}
\end{figure}

We characterized both theoretically and experimentally the average flow rate as a function of the contraction frequency (1/T) for symmetric contraction waves (Extended Data Fig. \ref{FigSI2}). While the deformation-imposed model (Fig. \ref{Fig1} of the Main text and Supplementary information Section 3) predicts linear scaling with the frequency, we observe in the experiments performed on the soft vessel a saturation of the average flow rate at contraction frequencies larger than \SI{1}{\hertz}. The saturation is captured by the model considering an soft vessel and originates from incomplete relaxation of the soft vessel, which diminishes the effective actuation amplitude and thereby drives the transition from linear to sublinear scaling (Extended Data  Fig. \ref{FigSI2}, continuous line). 

%This saturation is explained captured by our model (\ref{FigSI2} continuous line and Supplementary information). 

\begin{figure}[t!]
%\hspace{-2.2cm}
\centering
\hspace{0in}
\includegraphics[width=0.5\textwidth,trim=0cm 0cm 0cm 0cm]{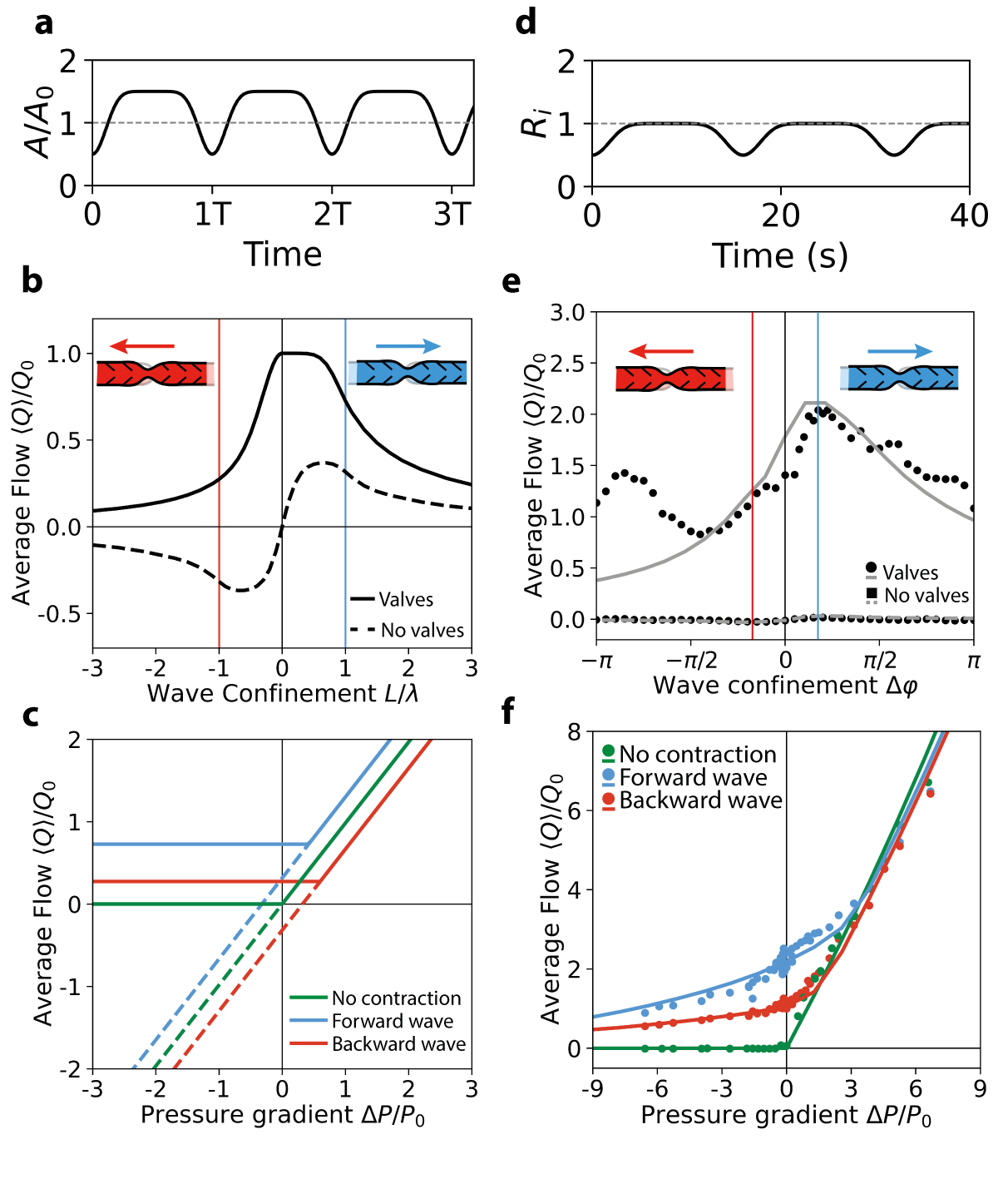}
\caption{\linespread{1.1}\selectfont{}
\textbf{Scalable and robust flow rectification for pulsating waves.} (\textbf{a,b,c}) For pulsatile contractions with an asymmetry towards relaxation \textbf{(a)}, optimal transport is observed for forward propagating waves \textbf{(b,c)}. This observation is confirmed experimentally and theoretically for soft vessels (\textbf{d,e,f}).
}
\label{FigSI1}
\end{figure}

We extended our study to pulsating waves characterized by a longer relaxation phase (Extended Data Fig. \ref{FigSI1}). Both theoretical and experimental results show that forward propagating waves increase fluid transport compared to backward-propagating waves.

\bibliography{bibliography.bib}

\newpage

\newpage

% \begin{table}[]
%     \centering
%     \begin{tabular}{|c|c|}
%     \hline
%     Physical parameter & Experimental value \textcolor{red}{JP's measurement} \\
%     \hline \hline
%          Inner radius $R_0$ & 0.5 cm \textcolor{red}{\SI{0.55}{\centi\meter}} \\
%          Tube thickness $h$ & 0.5 cm \textcolor{red}{\SI{0.3}{\centi\meter}} \\
%          Channel length $L$ & 30 cm \textcolor{red}{\SI{42}{\centi\meter} (6 valves) \SI{62}{\centi\meter} (9 valves)} \\
%          Tube Young's modulus $E$ & 0.5$\times 10^6$ Pa \\
%          Tube Poisson ratio $\nu$ & 0.45\\
%          Fluid density $\rho$ & $1000$ kg/m$^3$ \textcolor{red}{\SI{1060}{\kilo\gram\per\cubic\meter}}\\
%          Fluid viscosity $\mu(\SI{15}{\celsius})$ & \SI{195}{\milli\pascal\second}\\
%          Fluid viscosity $\mu(\SI{20}{\celsius})$ & \SI{175}{\milli\pascal\second}\\
%          Fluid viscosity $\mu(\SI{25}{\celsius})$ & \SI{135}{\milli\pascal\second}\\
%          Open valve resistance $\mathcal{R}_v$ & $1.5 \cross 10^{10}$ Pa s / m$^3$\\
%          Elastic modulus $R_0 \frac{\partial P}{\partial R}$ &  $1.9 \cross 10^4$ Pa\\
%          Hydraulic resistances $[\SI{}{\milli\bar\second\per\cubic\meter}]$ & pump 1.443 \\
%          & channel 2.315 \\
%          & connections 1.218 \\
         
%     % Rp = 1.443 # [mbar.s/m3] sole pump
%     % R0 = 3.533 # [mbar.s/m3] channel + connections
%     % Rc = 1.218 # [mbar.s/m3] connections
%     % R0b = 2.315 # [mbar.s/m3] channel - connections 
    
%     \hline
%     \end{tabular}
%     \caption{Experimental parameters}
%     \label{tab:my_label}
% \end{table}

\end{document}